\documentstyle[11pt,newpasp,twoside,epsf]{article}
\def\edcomment#1{\iffalse\marginpar{\raggedright\sl#1\/}\else\relax\fi}
\reversemarginpar

\begin{document}
\title{NUCLEAR SPIRALS AND SUPERMASSIVE BLACK HOLES}
\author{H. B. Ann and Parijat Thakur}
\affil{Division of Science Education, Pusan National University, Busan 609-
735, Korea}

\begin{abstract}
We have performed SPH simulations for the response of the  gaseous disks to  
the imposed potentials including those from bars and SMBHs.
Evolution of the nuclear regions of gaseous disks depends critically
on the masses of SMBHs as well as the sound speeds in the gas. 
\end{abstract}

\section{Introduction}

Nuclear spirals are known to be preponderant in active galaxies (Reagan \&
Mulchaey 1999; Martini et al. 2003). They have a variety of morphology,
from the grand-design symmetric two-armed spirals to chaotic ones (Martini 
et al. 2003). The flocculent nuclear spirals are considered to be formed by
the acoustic instability proposed by Montenegro, Yuan, \& Elmegreen (1998),
whereas the grand-design nuclear spirals are thought to be formed by
the hydrodynamical instability caused by the gas inflow driven by the
non-axisymmetric potentials (Englmaier \& Shlosman 2000).

It is well known that the response of gaseous disk to the imposed
non-axisymmetric potentials depends not only on the potential shape of the
model galaxy but also on the hydrodynamic properties of the gaseous
disk (Ann \& Lee 2000; Englmaier \& Shlosman 2000; Maciejewski et al. 2002).
However, the effects of SMBHs on the gas flow inside the ILRs have not been 
studied much. Here, we present some results of numerical experiments including 
SMBH for the
formation of nuclear spirals, based on smoothed particle hydrodynamics (SPH). 

\section{Models and Numerical Methods}

We have assumed that a barred galaxy is made up of three stellar
components (bulge, disk, bar) and two dark ones (SMBH, halo). We adopt
simple analytic forms for the potential generated by each component.
The properties of all the potential generating
components are assumed to be invariant in time. 
We considered mass models which are thought to resemble early type 
galaxies ($\sim$ SBa) by assuming the bulge-to-disk mass
ratios as $0.5$. We assumed a strong bar which
has the fractional mass of 0.2 and the axial ratio (a/b) as 3. 
The bar rotation period is $1.4\times 10^8$ years. 
We used isothermal assumption for the
gas but we explored the effect of gas temperature by varying the sound speed
of gas. We adopted  $\alpha =1.0$ and $\beta=2.5$ for the artificial viscosity 
coefficients. The self-gravity of gas was also included.

\section{Results and Discussion} 

Fig. 1 shows snapshots of the evolution of the nuclear regions of gaseous
disks at the evolution time of 20 bar rotations. The three models have 
the same mass distributions except for the central SMBHs. The model M1 has
no SMBH, while the other two models (M2, M3) have a SMBH whose mass is
about $1\%$ of the total mass of the visible components (disk, bulge and bar)
but they assumed different sound speeds in the gas; 10 km/s for M2 and
15 km/s for M3. The sound speed in the gas of M1 model
is the same as that of the model M2. 
 
As shown clearly in Fig. 1, the nuclear regions of the gaseous disk of
M1 model evolves to leading spirals between the IILR and OILR,
whereas those of M2 and M3 develop
trailing spirals whose detailed shapes depend on the sound speeds in 
the gas. The cold gaseous disk assumed in the model M2 shows ring-like 
spirals, while the hot gaseous disk of the model M3 shows tightly wound 
spirals whose innermost parts reach close to the center. 
Thus, it seems quite clear that the tightly wound trailing nuclear spirals can 
be developed in the hot interstellar medium when there is a SMBH whose mass is
large enough to remove the IILR. This is the reason why nuclear spirals are
frequently observed in active galaxies.

\begin{figure}
\plotone{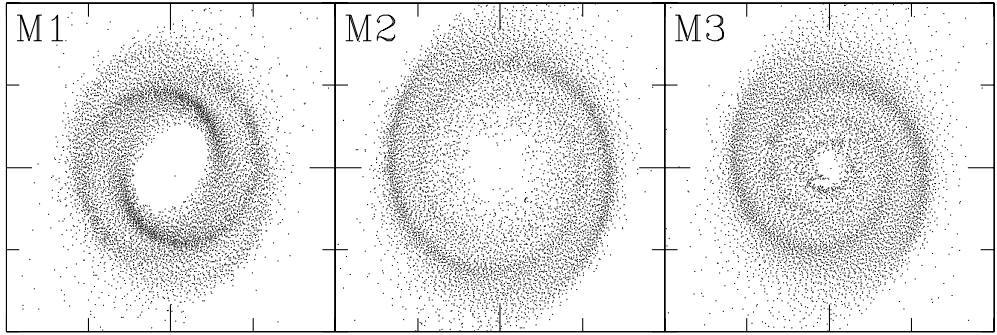}
\caption{Snapshots of the evolution of gaseous disks at 20 bar rotations. 
The box size is 2 kpc in one dimension.
}
\end{figure}

\acknowledgments
This work has been supported in part by the ARCSEC. 
The computations are conducted using the facilities
of KISTI.

\end{document}